\documentclass{appolb}
\usepackage{graphicx}
\begin{document}
\title{The nuclear symmetry energy and other isovector 
observables from the point of view of nuclear structure%
%\thanks{Presented at }%
}
\author{G. Col\`o, X. Roca-Maza
\address{Dipartimento di Fisica, Universit\`a degli Studi di Milano, \\
and INFN, Sezione di Milano, Via Celoria 16, 
Milano (Italy)}
\vspace{0.3cm}
\\ N. Paar
\address{Physics Department, Faculty of Science, 
University of Zagreb, Zagreb (Croatia)}
}
\maketitle
\begin{abstract}
In this contribution, we review some works related
with the extraction of the symmetry energy parameters from
isovector nuclear excitations, like the giant resonances.
Then, we move to the general issue of how to assess whether correlations between
a parameter of the nuclear equation of state and a nuclear
observable are robust or not. To this aim, we introduce the
covariance analysis and we discuss some counter-intuitive, yet
enlightening, results from it.
\end{abstract}
%\PACS{...}
  
\section{Introduction}

Among the widely debated questions in nuclear physics, we can
mention the one related to the nuclear equation of state and its extrapolation
to extreme conditions. Even restricting to zero temperature, the nuclear
community is still striving to determine the behaviour of the
energy per particle in uniform matter as a function of density and neutron-proton
asymmetry. The energy as a function of the neutron-proton asymmetry or,
in turn, the so-called symmetry energy (cf. below for a precise
defintion) is of particular interest because of its impact on
the physics of exotic, neutron-rich or neutron-deficient, nuclei.
It also affects in an important fashion the properties of some astrophysical compact objects
like the neutron stars. Topical conferences address the problem of the
determination of the nuclear symmetry energy, and the reader can 
consult a recent topical volume to check the present status of
our understanding \cite{topical}.

One can expect that the study of isovector modes of finite nuclei can shed light
on the problem of the determination of the symmetry energy and of
its density dependence. In the isovector collective motion, protons
are displaced with respect to neutrons: in other words, one creates
locally a neutron-proton imbalance. Thus, the response of the nucleus to this
perturbation is related to the variation of the energy density as a function
of asymmetry. However, the nucleus is different from uniform matter: 
not only shell effects, or pairing, are expected to play a role, 
but isospin is not a good quantum number 
and the separation of isoscalar and isovector motion raises some concern.
We briefly discuss some works that are related with the constraints on
the symmetry energy emerging from the analysis of isovector properties of
finite nuclei, and we stress the consistency of the results we have obtained. 

Then, we introduce covariance analysis as a rigorous way to
determine whether extracting constraints on the symmetry energy 
from isovector observables is justified or not. Until recently,
covariance analysis has not been object of much interest by nuclear
theorists. We introduce its basic concepts and try to make the reader
familiar with the idea that there is a quantitative fashion to
determine whether two quantities $A$ and $B$ are correlated
or not, whenever calculated in a given framework. In the present context,
$A$ and $B$ could be, respectively, a parameter characterizing
the density behaviour of the symmetry energy and an isovector observable.
Our discussion will be nevertheless more general.
Our analysis is based on Energy Density Functional (EDF) 
calculations \cite{Bender:2003}.
We will show that the correlations that emerge are not universal
but {\em will, to some extent, depend on the chosen model.}

\section{General definitons: the nuclear equation of state and
the symmetry energy}

We assume that the nuclear systems, as any Fermi system, can be
described in terms of a {\em local} energy functional. This implies
that we can write their total energy as 
\begin{equation}\label{edf}
E = \int d^3r\ {\cal E}(\rho_n(\vec r),\rho_p(\vec r)),
\end{equation}
where $\cal E$ is the energy density and $\rho_n$ and $\rho_p$ are,
respectively, neutron and proton densities. Instead of $\rho_n$ and 
$\rho_p$, one can use the total density $\rho$ and the {\em local} 
neutron-proton asymmetry,
\begin{equation}
\beta \equiv {\rho_n-\rho_p \over \rho}.
\end{equation}
In asymmetric matter, we can make a Taylor expansion of 
${\cal E}(\rho,\beta)$ in $\beta$ and retain only the quadratic term 
(odd powers of $\beta$ are forbidden due to isospin symmetry), 
\begin{eqnarray}\label{def_sym}
{\cal E}(\rho,\beta) & \approx & {\cal E}_0(\rho,\beta=0) +
{\cal E}_{\rm sym}(\rho) \beta^2 \nonumber \\
& = & {\cal E}_0(\rho,\beta=0) +
\rho S(\rho) \beta^2.
\end{eqnarray}
The first term on the r.h.s. is the energy density of symmetric nuclear 
matter while the second term defines the the symmetry 
energy $S(\rho)$. The expansion (\ref{def_sym}) should continue with
a quartic and possibly with higher order terms in $\beta$; however,
the coefficient of the term in $\beta^4$ is, to the best of our 
knowledge, found to be negligible in most models at the densities 
of interest for this work \cite{vidana2009,Trippa:2008}.

If we focus on the density dependence of $S$ close to the usual 
nuclear density we can define
\begin{eqnarray}\label{parameters}
J & \equiv & S(\rho_0), \nonumber \\
L & \equiv & 3\rho_0\ S^\prime(\rho_0), 
\end{eqnarray}
where $\rho_0$ is the saturation density for symmetric nuclear matter,
$\rho_0 \approx$ 0.16 fm$^{-3}$. $L$ is often referred to as the 
``slope parameter''. 

We shall, in what follows, discuss how these parameters
can be extracted from the comparison between theoretical calculations
and experimental measurements of the properties of isovector states. The
theoretical calculations are done using the Hartree-Fock (HF) plus
Random Phase Approximation (RPA) framework. We do not provide here
information about this well-known scheme. The details of our
implementation, together with a general introduction, can be 
found in Ref. \cite{cpc}.

\section{Symmetry energy extracted from giant dipole, giant 
quadrupole and pygmy dipole resonances}

\subsection{IVGDR}

The case of the most collective and well known isovector giant 
resonance, namely the isovector giant dipole resonance (IVGDR), 
has been studied in Ref.~\cite{Trippa:2008}. In that work, the starting
point is the hydrodynamical model proposed by E. Lipparini and S. 
Stringari~\cite{Lipparini:1989}. If one denotes by $m_k$ the $k$-th moment 
associated with the strength function of an  external operator $F$, the
IVGDR energy can be evaluated within the framework of the hydrodinamical model
if defined as $E_{-1}\equiv \sqrt{m_1/m_{-1}}$. The result is 
\begin{equation}\label{Els}
E_{-1}\equiv\sqrt{\frac{m_1}{m_{-1}}}
=\sqrt{\frac{3\hbar^2}{m\langle
r^2\rangle}\frac{b_{\rm vol}}{\left(1+\frac{5}{3}\frac{b_{\rm surf}}
{b_{\rm vol}}A^{-\frac{1}{3}} \right)}(1+\kappa)},
\end{equation}
where $b_{\rm vol}$ and $b_{\rm surf}$ are the volume and surface 
coefficients of the macroscopic symmetry energy and  $\kappa$ is the well-known 
``enhancement factor'', which in the case of Skyrme forces is associated with 
their velocity dependence \cite{Bender:2003}. The coefficient $b_{\rm vol}$ 
can be identified with $S(\rho_0)\equiv J$; if the nucleus had a sharp surface this would be 
the only quantity appearing in the previous expression. The nuclear surface does
manifest itself in the correction $\left( 1+\frac{5}{3}\frac{b_{\rm surf}}
{b_{\rm vol}}A^{-\frac{1}{3}} \right)^{-1}$. To connect this with the 
microscopic symmetry energy $S$, one can rewrite this correction and assume  
that the r.h.s. of Eq. (\ref{Els}) in a heavy nucleus does not scale as $\sqrt{S(\rho_0)}$, 
but rather as $\sqrt{S(\bar \rho)}$ where $\bar \rho$ is some value of density below 
the saturation density  $\rho_0$. In Ref. \cite{Trippa:2008} it has been 
found that such a correlation between $E_{-1}$ (calculated within HF-RPA) 
and $\sqrt{S(\bar \rho)}$ exists, with $\bar \rho$ around 0.1 fm$^{-3}$ for
heavy nuclei (in agreement with \cite{prl}). By exploiting this correlation 
%in the form
%\begin{equation}\label{fit1}
%\sqrt{S(\bar \rho)(1+\kappa)} = a + b E_{-1}(RPA),
%\end{equation}
and inserting the experimental value for the IVGDR energy in $^{208}$Pb, 
it has been found that  
\begin{equation}\label{finalcon}
23.3\;{\rm MeV} < S(0.1) <24.9\;{\rm MeV}.
\end{equation}

\subsection{IVGQR}

It is a natural question to ask whether the isovector giant quadrupole resonance 
(IVGQR) provides a consistent extraction of the density dependence of the 
symmetry energy. Until recently,
the experimental properties of the IVGQR had not been determined very accurately;
this goal has been achieved using a very intense and polarized photon beam at the
HI$\vec\gamma$S facility \cite{Henshaw:2011}. In Ref. \cite{Roca-Maza:2013a},
a comprehensive theoretical analysis has been performed, based both on RPA
calculations and on a macroscopic interpretation. The energy of the IVGQR receives 
contribution from unperturbed particle-hole (p-h) configurations at 2$\hbar\omega$ 
excitation energy, plus some correlation energy related to the isovector 
residual interaction. This idea has been implemented in Ref.
\cite{Roca-Maza:2013a}, with mild assumptions and taking care
of the fact that the unperturbed p-h energy can be related to the effective mass
and, in turn, to the isoscalar GQR energy. The main result is
\begin{equation}
E_{\rm IVGQR} \approx 2 \left[\frac{\left(E_{\rm ISGQR}\right)^2}{2}+ 
2\frac{\varepsilon_{{\rm F}_\infty}^2}{A^{2/3}}
\left(\frac{3S(\bar\rho)}{\varepsilon_{{\rm F}_\infty}} - 
1\right)\right]^{1/2},
\label{ex-ivgqr-3}
\end{equation}
where $\varepsilon_{{\rm F}_\infty}$ is the Fermi energy for
symmetric nuclear matter at saturation density, and $S(\bar\rho)$ is the symmetry energy
at some average nuclear density. If we take for this the same value that we have
adopted in the above discussion for the IVGDR, that is, 0.1 fm$^{-3}$, 
we can reproduce the experimental IVGQR energy; to turn it around, from
the two experimental IVGDR and IVGQR energies we can derive consistent values
for the value of $S(0.1)$. 

In Ref. \cite{Roca-Maza:2013a} it has been checked, in addition, that both
Skyrme and relativistic mean-field models follow quite well the
scaling predicted by Eq. (\ref{ex-ivgqr-3}). In fact, new Skyrme interactions 
have been fitted in Ref. \cite{Roca-Maza:2013a} using the same protocol as in 
Ref. \cite{Roca-Maza:2012b}, where the new set SAMi has been introduced; moreover, 
new effective Lagrangians have been fitted along the line of the DD-ME
one \cite{ddme}. These sets are characterized by different values of $J$ and
of the effective mass $m^*/m$. The microscopic results obtained with sets
having the same value of $m^*/m$ have been used as follows. The IVGQR energy 
can be reproduced only with forces having a specific combination of $J$ and $L$;
if we assume a value of $J$ = 32$\pm$1 MeV we extract
\begin{eqnarray}
L & = & 37 \pm 18\ {\rm MeV}; \label{l-gqr}\\
\Delta R \left( ^{208}{\rm Pb} \right) & = & 0.14\pm0.03\ {\rm fm}. \label{dr-gqr} 
\end{eqnarray} 
The second line corresponds to the neutron skin in $^{208}$Pb, that is
well known to be correlated with $L$ \cite{Brown:2000}. 

\subsection{PDR}

Among isovector modes, the so-called ``Pygmy Dipole Resonance'' (PDR) has
captured noticeable interest in the last decade. This definition is not free
from ambiguities if used to label generically the dipole strength below
the IVGDR. 
%In light, neutron-rich halo nuclei this strength may be
%specially enhanced due to the large transition probability of weakly bound
%neutrons to continuum states (``threshold effect''). In medium-heavy (even 
%unstable) nuclei, the neutron excess gives rise only to a neutron skin and not
%to a halo. In such systems a possible, peculiar mode of vibration has been proposed, 
%namely the oscillation of the neutrons of the skin with respect to
%the (essentially N$\approx$Z) core. The frequency of this mode should be
%lower than the IVGDR and the name PDR may be employed. However, very often
%this simple picture does not seem to hold because there the strength below
%the IVGDR seems to be fragmented in small peaks. 
Experimentally, low-lying 
dipole strength has been found in several nuclei \cite{Paar:2007,Zilges:2013}. 
Typically, the PDR strength may arrive up to a few \% of the dipole EWSR.

In Ref. \cite{Carbone:2010}, it has been proposed that the 
fraction of EWSR exhausted by the PDR and the slope parameter 
$L$ defined in Eq. (\ref{parameters}) are correlated.
This correlation could be explained if PDR is a mode related to
the oscillation of the excess neutrons, whose dynamics is decoupled from the 
IVGDR (see Ref. \cite{Suzuki:1990} for a transparent interpretation).
However, this picture may break in some cases. The collectivity of the
PDR seems to be somewhat model-dependent, and the states in that energy 
region have also a mixed isovector/isoscalar character \cite{Nazarewicz:2010,
Roca-Maza:2012a,Vretenar:2012}. A toroidal component has been found in
the calculations of
Refs. \cite{Vretenar:2001,Repko:2013}. Despite these well-taken warnings,
the values of the slope parameter and neutron skin (for $^{208}$Pb) that
have been extracted in \cite{Carbone:2010}, namely
\begin{eqnarray}
L & = & 64.8 \pm 15.7\ {\rm MeV}; \label{l-pdr}\\
\Delta R \left( ^{208}{\rm Pb} \right) & = & 0.194\pm0.024\ {\rm fm}, 
\label{dr-pdr} 
\end{eqnarray}
are consistent with the previous values given in Eq. (\ref{l-gqr}) and
(\ref{dr-gqr}). 

\subsection{Total dipole polarizability}

The total dipole polarizabilty $\alpha_D$ is proportional to the
inverse-energy weighted sum rule $m_{-1}$ of the dipole operator,
the exact relationship being
\begin{equation}
\alpha_D = \frac{8\pi e^2}{9}m_{-1}.
\end{equation}
The correlations between this quantity, calculated by means of a large
bunch of EDFs, and either the slope parameter $L$ or the neutron skin 
has been analyzed in Ref. \cite{Piekarewicz:2012}. In that work,
however, the correlation shows up very clearly only within families
made up with similar models. In fact, with the help of the droplet
model (DM), it has been shown in Ref. \cite{Roca-Maza:2013b} that
the slope parameter $L$ is well correlated with the product of $J$ times 
the dipole polarizability. The formula suggested by the DM is
\begin{equation}
\alpha_D^{\rm DM} \approx 
\frac{\pi e^{2}}{54} \frac{A \langle r^2\rangle}{J}\left[1+\frac{5}{2}
\frac{\Delta r_{np}^{\rm DM}+ \sqrt{\frac{3}{5}}\frac{e^2Z}{70 J} - 
\Delta r_{np}^{\rm surf}}{\langle r^2\rangle^{1/2}(I-I_C)}\right], 
\end{equation}
where the quantities appearing in the second term within square brackets
are defined in \cite{Roca-Maza:2013b} and shown to not vary appreciably
among models (at least in heavy systems like $^{208}$Pb \cite{prc}).  
Microscopic calculations obey such kind of scaling
quite well and allow extracting, assuming $J = 31\pm 2$ MeV, the value
\begin{equation}
L = 43 \pm 16\ {\rm MeV}. \label{l-dpol}
\end{equation}

\section{Covariance analysis}

All the above discussion is simply based on the empirical appearance
of a linear correlation between two observables when they are
calculated using several models like EDF parameterizations. 
This kind of analysis is not based on any statistical assumption
(models are assumed to be independent) but could be justified if
the correlation under study is suggested by macroscopic models.
A different strategy to judge about correlations between observables is
to use covariance analysis. The advantage is that there is
a more rigorous statistical foundation for such a method; 
however, this can be used only to judge whether
correlations exist {\em within the framework of a given model, 
whose parameters are varied without changing the
ansatz or the fitting protocol of such parameters.}
Although this will not be our focus here, covariance analysis 
is interesting for several more reasons. It allows estimating
the theoretical errors on the model parameters and thus, decide
if some of them is underconstrained or even redundant.

For the sake of brevity we do not give here a thorough explanation
of the method of covariance analysis. We refer to our recent
work \cite{cov} for better explanations, reminding also that 
many textbooks contain a more exhaustive treatment of the formalism 
(see e.g. \cite{Bevington}), and that an excellent introduction
for nuclear theorists is given in Ref. \cite{dobaczewski14}. 

\subsection{Summary of relevant formulas}

Let us consider a model, like an EDF one, characterized by $n$ 
parameters ${\bf p} = (p_1 , . . . , p_n)$. Observables ($A$) 
are functions of these parameters. Usually one builds the optimal
model, in the space defined by the $n$ parameters, by
$\chi^2$ minimization. 
%\begin{equation}
%\chi^2({\bf p}) = \sum_{\imath=1}^{m}
% \left(\frac{\mathcal{O}_\imath^{\rm theo.}({\bf p})-\mathcal{O}_\imath^{\rm ref.}}
% {\Delta\mathcal{O}_\imath^{\rm ref.}}\right)^2 \,
%\label{chi2}
%\end{equation}
%
%where ``theo.'' stands for the calculated values, and ``ref.'' may refer to experimental, 
%observational and/or {\it pseudo-data} that sometimes are used to guide the models. 
%Strictly speaking, $\Delta\mathcal{O}_\imath^{\rm ref.}$ should stand for the experimental standard deviations. 
%{\bf This choice is not always reasonable as in some cases the experimental error may be smaller than
%the accuracy of the fitted functional, and a small $\Delta\mathcal{O}_\imath^{\rm ref.}$ may prevent the fitting
%protocol from converging.}
Let us also assume that the $\chi^2$ is a well behaved, analytical 
hyper-function of the parameters around their optimal value ${\bf p}_0$, 
%\begin{equation}
%\partial_{{\bf p}}\chi^2({\bf p})\mid_{{\bf p}={\bf p}_0} = 0 \ \ ,
%\label{dchi2}
%\end{equation}
and that it can be approximated there by a Taylor expansion, namely
\begin{eqnarray}
\chi^2({\bf p})-\chi^2({\bf p}_0) &\approx& 
 \frac{1}{2}\sum_{\imath, \jmath}^n(p_{\imath}-p_{0\imath})
 \partial_{p_\imath}\partial_{p_\jmath}\chi^2(p_{\jmath}-p_{0\jmath})\nonumber\\
&\equiv& \sum_{\imath, \jmath}^n(p_{\imath}-p_{0\imath})
\left( \mathcal{E} \right)^{-1}_{\imath\jmath}(p_{\jmath}-p_{0\jmath}) \ \ ,
\label{curvature}
\end{eqnarray}
where we have defined the curvature matrix $\mathcal{E}^{-1}$ and
its inverse which is the covariance (or error) matrix $\mathcal{E}$. 
%From it, the correlation matrix ($\mathcal{C}$) can be estimated as  
%\begin{equation}
%\mathcal{C}_{\imath\jmath} \equiv \frac{\mathcal{E}_{\imath\jmath}}
% {\sqrt{\mathcal{E}_{\imath\imath}\mathcal{E}_{\jmath\jmath}}}.
%\label{c}
%\end{equation}
%$\mathcal{C}_{\imath\jmath}$ takes values form $-1$ (meaning 
%large anti-correlation between parameters $p_\imath$ and $p_\jmath$) 
%to $1$ (meaning large correlation).

Let us now expand an observable $A({\bf p})$ around the minimum 
${\bf p}_0$ assuming a smooth behavior and, therefore, neglecting second 
order derivatives:     
\begin{equation}
A({\bf p}) = A({\bf p}) + ({\bf p}-{\bf p}_0)
\partial_{{\bf p}}A({\bf p})\mid_{{\bf p}={\bf p}_0}
\equiv A_0 + ({\bf p}-{\bf p}_0){\bf A_0} \ \ .
\label{obs}
\end{equation}
The covariance between two observables is defined as 
\begin{equation}
C_{AB} = \overline{(A({\bf p})-\overline{A}) 
(B({\bf p})-\overline{B})}\ .
\label{co}
\end{equation}
Using the above expansion for the observables, and the fact that 
the parameters they depend upon are expected to have a Gaussian
distribution with curvature matrix $\mathcal{E}^{-1}$, one can write
\begin{equation}
C_{AB} \approx \sum_{\imath\jmath}^n 
\left.\frac{\partial A({\bf p})}{\partial p_\imath}\right 
\vert_{{\bf p}={\bf p}_0}
\mathcal{E}_{\imath\jmath}
\left.\frac{\partial B({\bf p})}{\partial p_\jmath}\right 
\vert_{{\bf p}={\bf p}_0}.
\label{co-2}
\end{equation}
The variance of $A$ is, then, given by $C_{AA}$. One may also calculate 
the Pearson-product moment correlation coefficient between those 
observables, that is,       
\begin{equation}
c_{AB} \equiv \frac{C_{AB}}{\sqrt{C_{AA}C_{BB}}} \ \ . 
\label{corr-pearson}
\end{equation}
%In analogy with the correlation coefficient defined in Eq.~(\ref{c}), 
$c_{AB}=1$ means complete correlation between observables $A$ and $B$, 
whereas $-1$ means complete anti-correlation and $c_{AB}=0$ 
means no correlation at all.

%uncomment the following lines to place a figure
\begin{figure}[htb]
\centerline{%
\includegraphics[width=8cm]{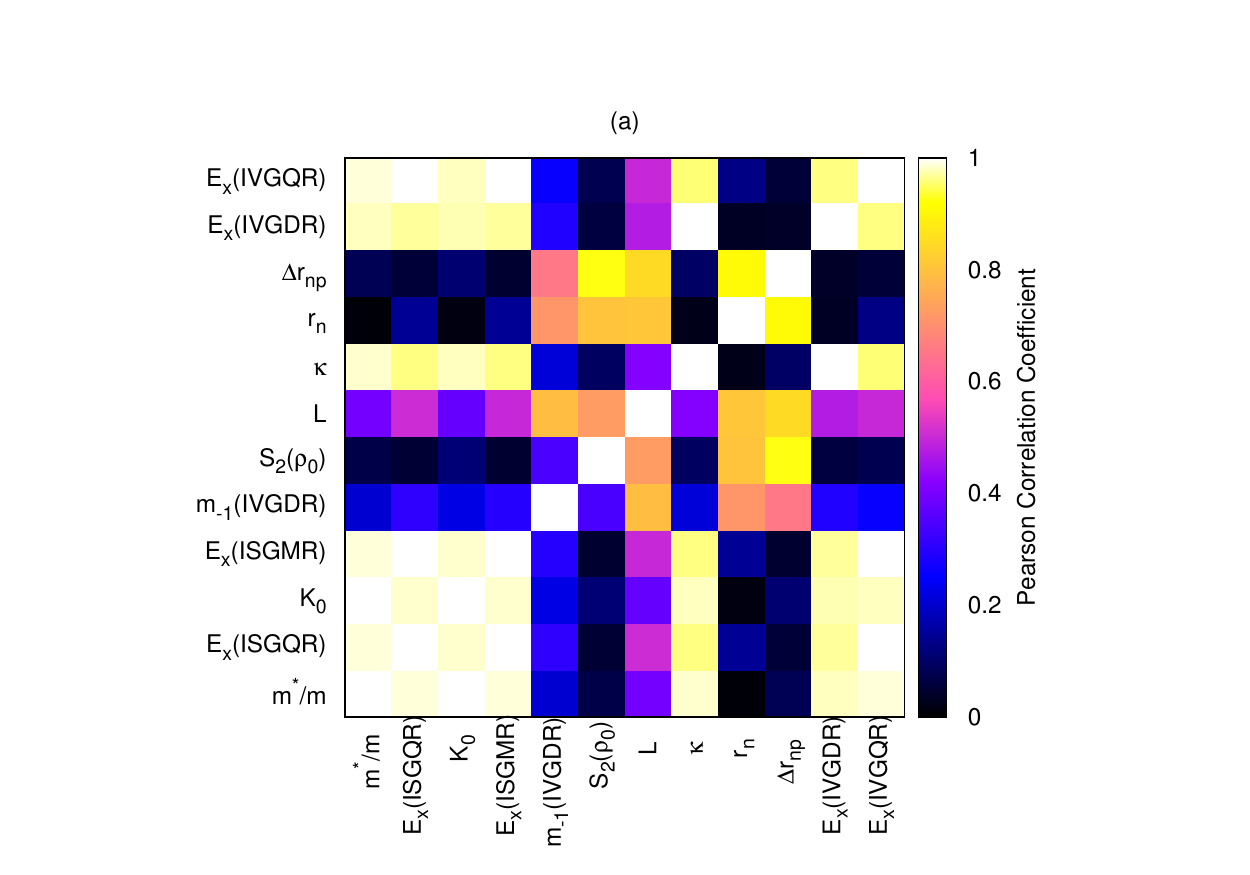}
\includegraphics[width=8cm]{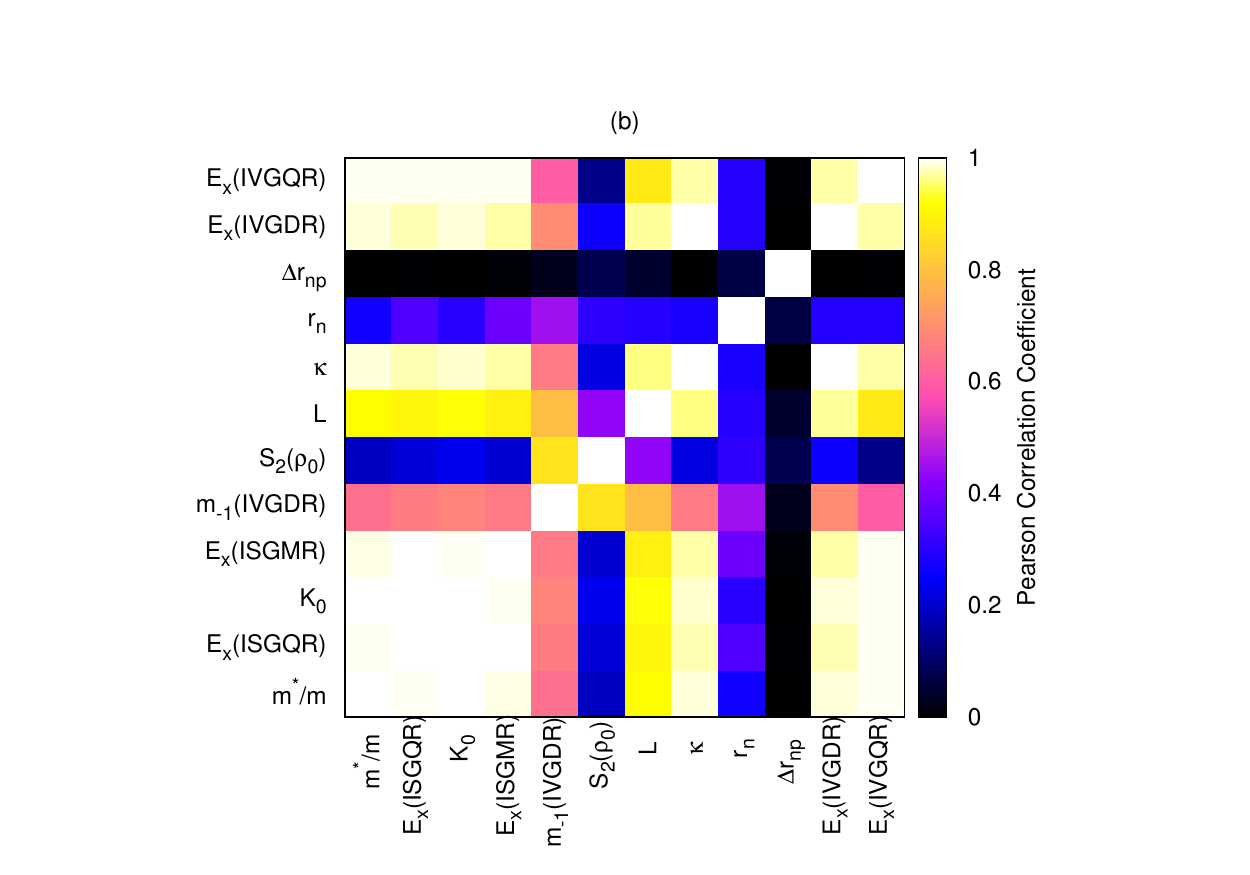}}
\caption{Values of the Pearson-product correlation coefficient 
calculated among pairs of observables. Panels (a) and (b) corresponds
to two different kinds of mimimization procedures, that are discussed
in the text.}
\label{fig1}
\end{figure}

\subsection{Some results from the covariance analysis}

Several results obtained via the covariance analysis have been
presented in Ref. \cite{cov}. Here, we only focus on some
illustrative findings that are of relevance in connection
with our previous discussion on symmetry energy and other
isovector observable.

We have fitted a functional called SLy5-min, that has been built 
so that to be as close as possible to the original SLy5 force
\cite{chabanat1997,chabanat1998}. Details and differences
are highlighted in Ref. \cite{cov}. We now illustrate what
happens to correlations between observables when one varies
the fitting protocol. 

The original SLy5 functional, and the present Sly5-min functional
as well, has been fitted by including in the $\chi^2$ as pseudo-data
a set of values for the neutron matter equation of state derived
by some {\em ab-initio} calculation with realistic forces 
\cite{Wiringa}. In the variant that we label as SLy5-a, we
have changed in the $\chi^2$ the weight associated with these pseudo-data, that is, 
with the equation of state of neutron matter. In particular, We have 
increased the value of the error on these points, $\Delta \frac{E}{A}(\rho,\delta=1)$, 
from $0.1\times \frac{E}{A}(\rho,\delta=1)$ -- that corresponds 
to a 10\% relative error -- to $0.5\times \frac{E}{A}(\rho,\delta=1)$. 
The Pearson-product correlation coefficients of this fit are shown in
panel (a) of Fig.~\ref{fig1}. One can notice a strong correlation between
the isovector observables, that are, the neutron radius of ${}^{208}$Pb, $S(\rho_0)$, 
$L$ and $m_{-1}$(IVGQR). This result clearly indicates that {\it when a constraint 
on a property $A$ is relaxed, correlations with other related observables $B$ not included 
in the fitting protocol become larger.}

The second variant we have built is denoted as SLy5-b. In this case, we have kept 
all terms in the $\chi^2$ as in SLy5-min except the equation of state of neutron 
matter that now is not included at all, 
and we added instead a very tight constraint on the 
neutron skin thickness of ${}^{208}$Pb: we have chosen 
$\Delta r_{np} = 0.160 \pm 0.001$ fm. 
Then, panel (b) of Fig.~\ref{fig1} shows another interesting outcome:
$\Delta r_{np}$ displays an almost vanishing correlation with all the other quantities.
This indicates that {\it when a property $A$ is tightly constrained in the fitting
protocol, correlations with other observables $B$ become very small.}       

\section{Conclusions}

The most important outcomes of our recent works on the constraints on
the density dependence of the symmetry energy imposed by the properties
of isovector excitations are
\begin{itemize}
\item if one looks at correlations between observables and values or
derivatives of the symmetry energy, calculated using a large set of 
different EDFs, consistent values of $L$ and of the neutron skin
thickness in $^{208}$Pb are extracted from the study of the IVGDR, 
IVGQR, PDR and total dipole polarizability [cf., in particular, Eqs.
(\ref{l-gqr}), (\ref{l-pdr}), (\ref{l-dpol}) as well as Eqs. 
(\ref{dr-gqr}), (\ref{dr-pdr})];
\item some of these correlations are suggested by macroscopic models
and, in general, we believe that it would be desirable that correlations
emerge from a sound physical picture rather than simply from 
a numerical analysis;
\item models may, or may not, display the correlations that one expects
on physical grounds because of the strong impact of the fitting protocols.
We have shown that correlations of an observable $A$ with related 
observables $B$ may emerge (vanish) if the constraint put on $A$ in the
fitting protocol is negligible (very tight).
\end{itemize}

\end{document}